\documentstyle[sprocl]{article}

\newcommand{\bea}{\begin{eqnarray}}\newcommand{\eea}{\end{eqnarray}}

\newcommand{\brr}{\begin{array}}\newcommand{\err}{\end{array}}
\newcommand{\bit}{\begin{itemize}}\newcommand{\eit}{\end{itemize}}
\newcommand{\ben}{\begin{enumerate}}\newcommand{\een}{\end{enumerate}}

\def\lan{\langle}
\def\lf{\left}

\def\non{\nonumber}
\def\pa{\partial}\def\ran{\rangle}
\def\rar{\rightarrow}
\def\ri{\right}

\def\al{\alpha}\def\bt{\beta}\def\ga{\gamma}
\def\de{\delta}\def\De{\Delta}
\def\te{\theta}

\def\si{\sigma}
\def\om{\omega}\def\Om{\Omega}
\def\CP{{_{CP}}}
\def\T{{_{T}}}

\def\1{{_{1}}}
\def\2{{_{2}}}
\begin{document}

\title{UNDERSTANDING FLAVOR MIXING \\[2mm] IN  QUANTUM FIELD THEORY}

\author{M. BLASONE}

\address{Blackett  Laboratory, Imperial  College,
\\
Prince Consort Road, London  SW7 2BZ, U.K. \\
E-mail: m.blasone@ic.ac.uk}

\author{A. CAPOLUPO, G. VITIELLO}

\address{Dipartimento di  Fisica and INFN,
\\
 Universit\`a di Salerno, I-84100 Salerno, Italy\\
E-mails: capolupo@sa.infn.it, vitiello@sa.infn.it}

\maketitle

\begin{abstract}
We report on recent results showing that
a rich non-perturbative vacuum structure is
associated with flavor mixing in Quantum Field Theory.
We treat explicitly the case of mixing among three generations of Dirac fermions.
Exact oscillation formulas are presented exhibiting new features with respect to
the usual ones. CP and T violation are also discussed.
\end{abstract}

\section{Introduction}
In recent years, there has been much progress in
the understanding of flavor mixing
in Quantum Field Theory (QFT).
The original discovery of the unitary inequivalence
of the mass and the flavor representations in QFT \cite{BV95}, has prompted
further investigations on fermion mixing
\cite{BHV99,remarks,currents,hannabus,fujii,fujii2} as well as on boson
mixing\cite{lathuile,binger,bosonmix,Ji2}.
It has emerged that the rich non--perturbative
vacuum structure associated with field mixing, leads to phenomenologically
relevant modification of the flavor oscillation formulas, exhibiting new
features with respect to the usual quantum--mechanical ones \cite{Pontec}.

In the following we discuss three flavor fermion mixing explicitly. A
discussion of mixing of boson fields can be found in Ref.\cite{bosonmix}.

\section{Three flavor fermion mixing}

Let us consider the
following Lagrangian density, describing three free Dirac fields with a
mixed mass term (in the following we will refer explicitly to
neutrinos):
\bea\label{lagemu} {\cal L}(x)\,=\,  {\bar \Psi_f}(x) \lf( i
\not\!\partial -
  M \ri) \Psi_f(x)\, ,
\eea
where $\Psi_f^T=(\nu_e,\nu_\mu,\nu_{\tau})$ and $M^\dag =M$.

Among the various possible parameterizations of the three fields mixing
matrix, we choose to work with the CKM matrix\cite{CKM}:
\bea\label{fermix}
\Psi_f(x)  =\lf(\begin{array}{ccc}
  c_{12}c_{13} & s_{12}c_{13} & s_{13}e^{-i\delta} \\
  -s_{12}c_{23}-c_{12}s_{23}s_{13}e^{i\delta} &
  c_{12}c_{23}-s_{12}s_{23}s_{13}e^{i\delta} & s_{23}c_{13} \\
  s_{12}s_{23}-c_{12}c_{23}s_{13}e^{i\delta} &
  -c_{12}s_{23}-s_{12}c_{23}s_{13}e^{i\delta} & c_{23}c_{13}
\end{array}\ri) \Psi_m(x)\; \; \;  \eea
with $c_{ij}=\cos\theta_{ij},  s_{ij}=\sin\theta_{ij}$ being
$\theta_{ij}$ the mixing angles, $\de$ the CP-violating phase and
$\Psi_m^T=(\nu_1,\nu_2,\nu_3)$.

Using Eq.(\ref{fermix}), we  diagonalize
Eq.(\ref{lagemu}) to the Lagrangian for three Dirac fields, with
definite masses:
\bea\label{lag12}
{\cal L}(x)\,=\,  {\bar \Psi_m}(x) \lf( i
\not\!\partial -  M_d\ri) \Psi_m(x)  \, ,
\eea
where $M_d = diag(m_1,m_2,m_3)$. To generate the mixing matrix
Eq.(\ref{fermix}), we define\cite{BV95}:
\bea\label{subgenerators}
&&G_{12}(t)= e^{\theta_{12}L_{12}(t)} \;\;,\;\;
G_{23}(t)=e^{\theta_{23}L_{23}(t)}\;\;,\;\;
G_{13}(t)=e^{\theta_{13}L_{13}(\delta,t)}\,,
\eea
where
\bea\non
L_{12}(t)&=&\int
d^{3}x(\nu_{1}^{\dag}(x)\nu_{2}(x)-\nu_{2}^{\dag}(x)\nu_{1}(x)),
\\ \label{L123}
L_{23}(t)&=&\int
d^{3}x(\nu_{2}^{\dag}(x)\nu_{3}(x)-\nu_{3}^{\dag}(x)\nu_{2}(x)),
\\ \non
L_{13}(\de,t)&=&\int
d^{3}x(\nu_{1}^{\dag}(x)\nu_{3}(x)e^{-i\delta}-\nu_{3}^{\dag}(x)
\nu_{1}(x)e^{i\delta}).
\eea
We can thus write Eq.(\ref{fermix}) in the form:
\bea\label{mix2}
&&\nu_{\si}^{\alpha}(x)=G^{-1}_{\bf \te}(t) \,
\nu_{i}^{\alpha}(x)\, G_{\bf \te}(t),
\eea
with $(\si,i)=(e,1), (\mu,2), (\tau,3)$ and
\bea\label{generator}
G_{\bf \te}(t)&\equiv &G_{23}(t)G_{13}(t)G_{12}(t).
\eea
From Eqs.(\ref{subgenerators})-(\ref{generator}), we see that
the phase $\delta$ is unavoidable for three fields mixing,
while it can be incorporated in the definition of the fields in the case
of two flavors. Note also that other parameterizations of the
mixing matrix can be reproduced by means of the generators
Eq.(\ref{subgenerators}), which do not commute among themselves (see below).

The free fields  $\nu_i$ can be quantized in the usual
way\cite{Itz} (we use $t\equiv x_0$):
\bea\label{freefields}
\nu_{i}(x) = \sum_{r} \int d^3 k \lf[u^{r}_{{\bf
k},i}(t) \al^{r}_{{\bf k},i}\,+ \,   v^{r}_{-{\bf k},i}(t)
\bt^{r\dag }_{-{\bf k},i}   \ri]
e^{i {\bf k}\cdot{\bf x}} ,\qquad i=1,2,3
\eea
with $\om_{k,i}=\sqrt{{\bf k}^2+m_i^2}$. The vacuum for
the mass eigenstates is denoted by $|0\rangle_{m}$:  $ \;
\al^{r}_{{\bf k},i}|0\rangle_{m}= \bt^{r }_{{\bf
k},i}|0\rangle_{m}=0$.   The anticommutation relations are the
usual ones; the wave function orthonormality and completeness
relations are those of Ref.\cite{BV95}.

By use of $G_{\bf \te}(t)$, the flavor fields can be expanded as:
\bea \non \label{mixfields1}
\nu_\si(x)&=& \sum_{r} \int d^3 k   \lf[ u^{r}_{{\bf k},i}(t)
\al^{r}_{{\bf k},\si}(t) +    v^{r}_{-{\bf k},i}(t)
\bt^{r\dag}_{-{\bf k},\si}(t) \ri]  e^{i {\bf k}\cdot{\bf x}}\,.
\eea
with $(\si,i)=(e,1) , (\mu,2) , (\tau,3)$ and $\al^{r}_{{\bf
k},\si}(t) \equiv G^{-1}_{\bf \te}(t)\al^{r}_{{\bf k},i}
G_{\bf \te}(t)$ and $\bt^{r\dag}_{{-\bf k},\si}(t)\equiv
 G^{-1}_{\bf \te}(t) \bt^{r\dag}_{{-\bf k},i}
G_{\bf \te}(t)$.

The crucial point\cite{BV95}, is that the generator of the field mixing
Eq.(\ref{fermix}) does not annihilate the vacuum for the free fields. We
are thus lead to define a new state, the {\em flavor vacuum}, in the
following way:
\bea
|0(t)\rangle_{f}\,\equiv\,G_{\te}^{-1}(t)|0\rangle_{m}
\;.
\eea
The unitary inequivalence (in the infinite volume limit) of $|0\rangle_{m}$
with $|0(t)\rangle_{f}$ has been rigorously
 proved for an arbitrary number of generations\cite{hannabus}.

The explicit form of the above defined flavor annihilation
operators (in the reference frame ${\bf k}=(0,0,|{\bf k}|)$) is
\bea \non
\alpha_{{\bf k},e}^{r}(t)&=&c_{12}c_{13}\,\alpha_{{\bf k},1}^{r}
+ s_{12}c_{13}\left(U^{{\bf k}*}_{12}\,\alpha_{{\bf k},2}^{r}
+\epsilon^{r}
V^{{\bf k}}_{12}\,\beta_{-{\bf k},2}^{r\dag}\right)
\\
&&+ e^{-i\delta}\,s_{13}\left(U^{{\bf k}*}_{13}\,
\alpha_{{\bf k},3}^{r} +\epsilon^{r}
V^{{\bf k}}_{13}\,\beta_{-{\bf k},3}^{r\dag}\right) ,
\\[2mm]\non
\alpha_{{\bf k},\mu}^{r}(t)&=&\left(c_{12}c_{23}- e^{i\delta}
\,s_{12}s_{23}s_{13}\right)\,\alpha_{{\bf k},2}^{r}
+\,s_{23}c_{13}\left(U^{{\bf k}*}_{23}\,\alpha_{{\bf k},3}^{r} +
\epsilon^{r} V^{{\bf k}}_{23}\,\beta_{-{\bf k},3}^{r\dag}\right)
\\
&& - \left(s_{12}c_{23}+e^{i\delta}\,c_{12}s_{23}s_{13}\right)
\left(U^{{\bf k}}_{12}\,\alpha_{{\bf k},1}^{r}
-\epsilon^{r} V^{{\bf k}}_{12}\,\beta_{-{\bf k},1}^{r\dag}\right)
\,,
\\[2mm]\non
\alpha_{{\bf k},\tau}^{r}(t)&=&c_{23}c_{13}\;\alpha_{{\bf k},3}^{r}
- \left(c_{12}s_{23}+e^{i\delta}\;s_{12}c_{23}s_{13}\right)
\left(U^{{\bf k}}_{23}\;\alpha_{{\bf k},2}^{r}
-\epsilon^{r} V^{{\bf k}}_{23}\;\beta_{-{\bf k},2}^{r\dag}\right)
\\
&&+\;\left(s_{12}s_{23}- e^{i\delta}\;c_{12}c_{23}s_{13}\right)
\left(U^{{\bf k}}_{13}\;\alpha_{{\bf k},1}^{r}
-\epsilon^{r} V^{{\bf k}}_{13}\;\beta_{-{\bf k},1}^{r\dag}\right)\;,
\eea

\pagebreak

\bea\non
\beta^{r}_{-{\bf k},e}(t)&=&c_{12}c_{13}\,\beta_{-{\bf k},1}^{r}
+ s_{12}c_{13}\left(U^{{\bf k}*}_{12}\,\beta_{-{\bf k},2}^{r}
-\epsilon^{r}V^{{\bf k}}_{12}\,\alpha_{{\bf k},2}^{r\dag}\right)
\\
&&+e^{i\delta}\, s_{13}\left(U^{{\bf k}*}_{13}\,\beta_{-{\bf k},3}^{r}
-\epsilon^{r} V_{13}^{{\bf k}}\,\alpha_{{\bf k},3}^{r\dag}\right)\,,
\\[2mm] \non
\beta^{r}_{-{\bf k},\mu}(t)&=&\left(c_{12}c_{23}- e^{-i\delta}\,
s_{12}s_{23}s_{13}\right)\,\beta_{-{\bf k},2}^{r}
+\, s_{23}c_{13}\left(U^{{\bf k}*}_{23}\,\beta_{-{\bf k},3}^{r} -
\epsilon^{r}\, V^{{\bf k}}_{23}\,\alpha_{{\bf k},3}^{r\dag}\right)
\\
&& -
\left(s_{12}c_{23}+e^{-i\delta}\,c_{12}s_{23}s_{13}\right)
\left(U^{{\bf k}}_{12}\;\beta_{-{\bf k},1}^{r} +\epsilon^{r}\;
V^{{\bf k}}_{12}\;\alpha_{{\bf k},1}^{r\dag}\right)\,,
\\[2mm] \non
\beta^{r}_{-{\bf k},\tau}(t)&=&c_{23}c_{13}\;\beta_{-{\bf k},3}^{r}
- \left(c_{12}s_{23}+e^{-i\delta}\;s_{12}c_{23}s_{13}\right)
\left(U^{{\bf k}}_{23}\;\beta_{-{\bf k},2}^{r} +
\epsilon^{r} V^{{\bf k}}_{23}\;\alpha_{{\bf k},2}^{r\dag}\right)
\\
&&+\;\left(s_{12}s_{23}- e^{-i\delta}\;c_{12}c_{23}s_{13}\right)
\left(U^{{\bf k}}_{13}\;\beta_{-{\bf k},1}^{r} +
\epsilon^{r} V^{{\bf k}}_{13}\;\alpha_{{\bf
k},1}^{r\dag}\right)\;.
\eea
with
\bea
&&V^{{\bf k}}_{ij}=|V^{{\bf
k}}_{ij}|\;e^{i(\omega_{k,j}+\omega_{k,i})t}\;\;\;\;,\;\;\;\;
U^{{\bf k}}_{ij}=|U^{{\bf k}}_{ij}|\;e^{i(\omega_{k,j}-\omega_{k,i})t}
\\ [2mm]
&&|U^{{\bf k}}_{ij}|=\cos \xi_{ij}^{{\bf
k}} \;\;\;\;,\;\;\;\;
|V^{{\bf k}}_{ij}|=\sin\xi_{ij}^{{\bf k}}
\\ [2mm]
&& \xi_{ij}^{{\bf k}}\,=\,\arctan\lf[\frac{k(\om_{k,i}-\om_{k,j}+
 m_{i}-m_{j})}{|k|^{2}+(\om_{k,i}+m_{i})(\om_{k,j}+m_{j})}\ri]
\eea
where $i,j=1,2,3$ and $j>i$.
The above operators annihilate $|0(t)\ran_f$ and
satisfy canonical anticommutation
relations (at equal times). Note that
\bea\label{relativ}
|U^{{\bf k}}_{ij}| \rar 1 \;\;\; ,\;\;\;|V^{{\bf k}}_{ij}|\rar 0\qquad
\;\;{\rm for}\;\; |k|\gg \sqrt{m_i m_j}\,,
\eea
%

\section{Currents and charges for three flavor fermion mixing}

Let us now consider the transformations acting on the triplet of free
fields with different masses. The Lagrangian Eq.(\ref{lag12})
 is invariant under global
$U(1)$ phase transformations  of the type $\Psi_m' \, =\, e^{i \al
}\, \Psi_m$: as a result, we have the conservation of the Noether
charge $Q=\int d^3x \, I^0(x) $, with $I^\mu(x)={\bar \Psi}_m(x)
\, \ga^\mu \, \Psi_m(x)$,  which is indeed the total charge of the
system (i.e. the total lepton number).

Consider now the $SU(3)$ transformations acting on $\Psi_m$:
\bea \label{masssu3} \Psi_m'(x) \, =\, e^{i \al_j  F_j}\,
\Psi_m (x) \, \qquad, \qquad
 j=1, 2,..., 8.
\eea
with $\al_j$ real constants, $F_j=\lambda_j/2$ and
$\lambda_j$ being the Gell-Mann matrices.
For $m_1\neq m_2\neq m_{3}$, the Lagrangian is not generally
invariant under (\ref{masssu3}) and  we obtain, by use of the
equations of motion,
\bea \non &&\de {\cal L}(x)\,= \,  i \al_j \,{\bar \Psi_m}(x)\,
[F_j,M_d ]\, \Psi_m(x) \, =\,  - \al_j \,\pa_\mu J_{m,j}^\mu
(x)
\\ [3mm]\label{fermacu1}
&&J^\mu_{m,j}(x)\, =\, {\bar \Psi_m}(x)\, \ga^\mu\,
F_j\, \Psi_m(x) \qquad, \qquad j=1, 2,..., 8. \eea
The explicit form of the  currents is:
\bea \non
&&J_{m,1}^\mu \, =\, \frac{1}{2} \lf[  {\bar \nu}_1 \ga^\mu \nu_2 \, +
\,{\bar \nu}_2 \ga^\mu \nu_1  \ri]
\quad ,\quad
J_{m,2}^\mu \, =\, -\frac{i}{2} \lf[  {\bar \nu}_1 \ga^\mu \nu_2 \, -
\,{\bar \nu}_2 \ga^\mu \nu_1  \ri]
\\
&&J_{m,3}^\mu \, =\, \frac{1}{2} \lf[  {\bar \nu}_1 \ga^\mu \nu_1 \, -
\,{\bar \nu}_2 \ga^\mu \nu_2  \ri]
\quad ,\quad
J_{m,4}^\mu \, =\, \frac{1}{2} \lf[  {\bar \nu}_1 \ga^\mu \nu_3 \, +
\,{\bar \nu}_3 \ga^\mu \nu_1  \ri]
\\ \non
&&J_{m,5}^\mu \, =\, -\frac{i}{2} \lf[  {\bar \nu}_1 \ga^\mu \nu_3 \, -
\,{\bar \nu}_3 \ga^\mu \nu_1  \ri]
\;\;\; ,\quad
J_{m,6}^\mu \, =\, \frac{1}{2} \lf[  {\bar \nu}_2 \ga^\mu \nu_3 \, +
\,{\bar \nu}_3 \ga^\mu \nu_2  \ri]
\\ \non
&&J_{m,7}^\mu  \,=\, -\frac{i}{2} \lf[  {\bar \nu}_2 \ga^\mu \nu_3 \, -
\,{\bar \nu}_3 \ga^\mu \nu_2  \ri]
\; \,,\, \;
J_{m,8}^\mu \, =\, \frac{1}{2\sqrt{3}} \lf[  {\bar \nu}_1 \ga^\mu \nu_1 \, +
\,{\bar \nu}_2 \ga^\mu \nu_2\,- 2{\bar \nu}_3 \ga^\mu \nu_3 \ri].
\eea
The charges  $Q_{m,j}(t)\equiv \int d^3 x \,J^0_{m,j}(x) $,
satisfy the  $SU(3)$  algebra at equal times:
$[Q_{m,j}(t), Q_{m,k}(t)]\, =\, i
\,f_{jkl}\,Q_{m,l}(t)$, with $f_{jkl}$ totally antisymmetric.
From (\ref{fermacu1}) we see  that
$Q_{m,3}$ and $Q_{m,8}$ are conserved as $M_d$ is diagonal.
We can define the combinations:
\bea
\non
Q_{1}& \equiv &\frac{1}{3}Q \,+ \,Q_{m,3}+
\,\frac{1}{\sqrt{3}}Q_{m,8},
\\ \label{noether1}
Q_{2}& \equiv & \frac{1}{3}Q \,- \,Q_{m,3}+\,\frac{1}{\sqrt{3}}Q_{m,8},
\\ \non
Q_{3}& \equiv &\frac{1}{3}Q \,- \,\frac{2}{\sqrt{3}}Q_{m,8},
\\ [2mm]
Q_i & = &\sum_{r} \int d^3 k\lf( \al^{r\dag}_{{\bf k},i}
\al^{r}_{{\bf k},i}\, -\, \bt^{r\dag}_{-{\bf k},i}\bt^{r}_{-{\bf
k},i}\ri)\quad ,\;\;\;  i=1, 2, 3 .
\eea
These are nothing but  the Noether charges associated with the
non-interacting fields $\nu_1$, $\nu_2$ and $\nu_3$: in the absence of
mixing, they are the flavor charges,  separately
conserved for each generation.
The generator of the mixing transformations can be now written
as:
\bea
G_{\bf \te}(t)\, =\, e^{i 2\te_{23} \,Q_{m,7}(t)}
\, e^{i 2\te_{13} \,Q_{m,5}(t)}
\, e^{i 2\te_{12} \,Q_{m,2}(t)}
\eea

Performing $SU(3)$ transformations on the flavor
triplet $\Psi_f$ gives a similar structure for the currents as before. The
related charges
$Q_{f,j}(t)$ $\equiv$  $\int d^3 x \,J^0_{f,j}(x) $ still close
the $SU(3)$ algebra. Due to the off--diagonal (mixing) terms in
the mass matrix $M$, $Q_{f,3}(t)$ and $Q_{f,8}(t)$ are time--dependent. This
implies an exchange of charge between $\nu_e$, $\nu_\mu$ and
$\nu_\tau$, resulting in the flavor oscillations.

In accordance with Eq.(\ref{noether1}), we now
define the {\em flavor charges} for mixed fields as
\bea\non
Q_e(t) & \equiv & \frac{1}{3}Q \, +
\, Q_{f,3}(t)\, + \,\frac{1}{\sqrt{3}} Q_{f,8}(t),
\\ \label{flavcha}
Q_\mu(t) & \equiv & \frac{1}{3}Q \, -
\, Q_{f,3}(t)+ \,\frac{1}{\sqrt{3}} Q_{f,8}(t),
\\ \non
Q_\tau(t) & \equiv & \frac{1}{3}Q \, -  \,
\frac{2}{\sqrt{3}} Q_{f,8}(t).
\eea
with $Q_e(t) \, + \,Q_\mu(t) \,+ \,Q_\tau(t) \, = \, Q$.
These charges have a simple expression in terms of the flavor
ladder operators ($\si= e,\mu,\tau$):
\bea Q_\si(t) & = & \sum_{r} \int d^3 k\lf( \al^{r\dag}_{{\bf
k},\si}(t) \al^{r}_{{\bf k},\si}(t)\, -\, \bt^{r\dag}_{-{\bf
k},\si}(t)\bt^{r}_{-{\bf k},\si}(t)\ri)\,, \eea
since they are connected to the Noether charges
$Q_i$  of Eq.(\ref{noether1}) via the mixing generator: $Q_\si(t)
= G^{-1}_\te(t)Q_i G_\te(t)$.

\section{Neutrino oscillations}

The oscillation formulas are
obtained by taking expectation values of the above charges on the
(flavor) neutrino state. Consider for example a $\rho$-flavor
neutrino state defined as $|\nu_\rho\ran \equiv \al_{{\bf k},\rho}^{r
\dag}(0) |0\ran_{f}$ (for a discussion on the  correct
definition of flavor states see Refs.\cite{BHV99,remarks,bosonmix}).
Working in the Heisenberg picture, we obtain:
\bea \non
{\cal Q}^\rho_{{\bf k},\si}(t)
&\equiv& \langle \nu_\rho|Q_\si(t)| \nu_\rho\rangle -
\, {}_f\lan 0 |Q_\si(t)| 0\ran_f
\\ [1.2mm]\label{charge1}
&=& \lf|\lf \{\al^{r}_{{\bf k},\si}(t), \al^{r \dag}_{{\bf
k},\rho}(0) \ri\}\ri|^{2} \,+ \,\lf|\lf\{\bt_{{-\bf k},\si}^{r
\dag}(t), \al^{r \dag}_{{\bf k},\rho}(0) \ri\}\ri|^{2}
\,,
\\ [3mm] \non
{\cal Q}^{\bar \rho}_{{\bf k},\si}(t)
&\equiv& \langle {\bar \nu}_\rho|Q_\si(t)| {\bar \nu}_\rho\rangle -
\, {}_f\lan 0 |Q_\si(t)| 0\ran_f
\\ [1.2mm]\label{charge2}
&=& - \lf|\lf \{\bt^{r}_{{\bf k},\si}(t), \bt^{r \dag}_{{\bf
k},\rho}(0) \ri\}\ri|^{2} \,- \,\lf|\lf\{\al_{{-\bf k},\si}^{r
\dag}(t), \bt^{r \dag}_{{\bf k},\rho}(0) \ri\}\ri|^{2}
\,,
\eea
where $|0\ran_{f}\equiv |0(0)\ran_{f}$ and
$|{\bar \nu}_\rho\ran \equiv \bt_{{\bf k},\rho}^{r
\dag}(0) |0\ran_{f}$. Charge
conservation is obviously ensured at any time:
$\sum_\si {\cal Q}_{{\bf k},\si}(t)\, = \, 1$.
We remark that the expectation value of $Q_\si$ cannot be
taken on vectors of the Fock space built on $|0\rangle_{m}$,
as shown in Refs.\cite{BHV99,remarks,bosonmix}.
Note that, in comparison
with the two-flavor mixing,
a vacuum contribution needs to be subtracted here:
due to the presence of the CP violating phase $\de$,
we have indeed  ${}_f\lan 0 |Q_\si(t)| 0\ran_f\neq 0$.

The oscillation  formulas for the flavor charges, on an
initial electron neutrino state, then follow:
\bea \non
&&{\cal Q}^e_{{\bf k},e}(t) \,
= \,1 - \sin^{2} 2 \te_{12} \cos^{4}\te_{13} \, \Big[
\cos^{2}\xi_{12}^{{\bf k}}
\sin^{2} (\De^{{\bf k}}_{12} t)
 +\,\sin^{2} \xi_{12}^{{\bf k}}   \sin^{2}
(\Om^{{\bf k}}_{12} t)\Big] \
\\  \label{chargee}
&& -\ \sin^{2} 2 \te_{13} \cos^{2} \te_{12}
\ \Big[\cos^{2} \xi_{13}^{{\bf k}}\
 \sin^{2} (\De^{{\bf k}}_{13} t)
 +\ \sin^{2} \xi_{13}^{{\bf k}} \  \sin^{2} (\Om^{{\bf k}}_{13} t)\Big] \
\\ \non && -\ \sin^{2} 2 \te_{13} \sin^{2} \te_{12}  \ \Big[
\cos^{2} (\xi_{12}^{{\bf k}}
-\xi_{13}^{{\bf k}})\  \sin^{2} (\De^{{\bf k}}_{23} t) +
\ \sin^{2}(\xi_{12}^{{\bf
k}}-\xi_{13}^{{\bf k}})\  \sin^{2}  (\Om^{{\bf k}}_{23} t)\Big],
\end{eqnarray}
and similar ones, with
$\De^{{\bf k}}_{ij}\equiv (\om_{k,j} -\om_{k,i})/2 $ and
$\Om^{{\bf k}}_{ij}\equiv (\om_{k,j} +\om_{k,i})/2 $.
These results are exact. The differences with respect to the usual
formulas for neutrino oscillations are in the energy dependence of
the amplitudes and in the additional oscillating terms.
In the relativistic limit of Eq.(\ref{relativ}),
the traditional QM (Pontecorvo) formulas are recovered.

\section{CP and T violation in QFT neutrino oscillations }

The CP violation developed in neutrino oscillations
is given in QM as:
\bea\label{CPQM}
\Delta_{\CP}(t)= P_ {\nu_{\sigma}\rightarrow \nu_{\rho}}(t) - P_
{\overline{\nu}_{\sigma}\rightarrow \overline{\nu}_{\rho}}(t).
\eea
where $ \sigma, \rho = e, \mu, \tau.$
The T violation can be obtained as:
\bea\label{TQM}
\Delta_{\T}(t)= P_ {\nu_{\sigma}\rightarrow \nu_{\rho}}(t) - P_
{\nu_{\rho}\rightarrow \nu_{\si}}(t). \eea
with $\Delta_{\CP}(t)= \Delta_{\T}(t)$
as a  consequence of CPT invariance.

The QFT analogue of Eq.(\ref{CPQM}) is
\bea\label{CPQFT}
\De^{\rho\si}_\CP(t) & \equiv &
{\cal Q}^\rho_\si(t)\,  +\, {\cal Q}^{\bar \rho}_\si(t)
\quad , \quad \rho,\si=e,\mu,\tau\,.
\eea
with
$\sum_\si \De^{e\si}_\CP(t) \, = \,0$ since
$ \sum_\si Q_\si(t) =Q $  and $\langle \nu_e|Q| \nu_e\rangle = 1 \, $,
$\langle {\bar \nu}_e|Q| {\bar \nu}_e\rangle =-1$.
For the case of an initial electron neutrino state, we obtain
\bea \non
\De^{e\mu}_\CP(t) &= & \frac{1}{2} \cos\te_{13} \,
\sin\de\,\sin(2\te_{12}) \, \sin(2\te_{13})\, \sin(2\te_{23})
\Big[\, U_{12}^2 \sin(2\De^{{\bf k}}_{12} t)
\\[2mm] \non
&&  -\,
V_{12}^2 \sin(2\Om^{{\bf k}}_{12} t)
 -U_{13}^2 \sin(2\De^{{\bf k}}_{13} t)  \, +\,
V_{13}^2 \sin(2\Om^{{\bf k}}_{13} t)
\\[2mm]\label{cpviol}
&& +(U_{12}^2 -  V_{13}^2)\sin(2\De^{{\bf k}}_{23} t)
\, +\, (V_{12}^2 -  V_{13}^2)
\sin(2\Om^{{\bf k}}_{23} t)\,\Big]
\eea
and $\De^{e\tau}_\CP(t) = - \De^{e\mu}_\CP(t)$.

For the study of T violation, defining  $\De_\T(t)$ as
$
\De^{\rho\si}_\T(t)  \equiv
\, {\cal Q}^\rho_\si(t)\,  -\, {\cal Q}^{\si}_\rho(t)
$
does not seem to work. We have indeed:
$\De^{\rho\si}_\T(t) \, - \,\De^{\rho\si}_\CP(t) \, \neq \, 0\,$
violating CPT invariance.
The correct definition is then:
\bea
 \De^{\rho\si}_\T(t)\, \equiv
\, {\cal Q}^\rho_\si(t)\,  -\, {\cal Q}^{\rho}_\si(-t)
\quad, \quad \rho,\si=e,\mu,\tau .
\eea
We obtain
$ \De^{e\si}_\T \, \neq \, 0$  for $\si\neq e$
and
$\De^{e\si}_\T \,=\,\De^{e\si}_\CP$
in agreement with
${\cal Q}^e_\si(-t)\, = \, - {\cal Q}^{\bar e}_\si(t)$.
Further work is in progress on this topic.

\section{Conclusions}

We have discussed the mixing of (Dirac) fermionic fields in Quantum Field
Theory for the case of three flavors.
We have constructed the flavor Hilbert space and studied the
currents and charges for mixed fields (neutrinos).

We have then derived the exact QFT oscillation formulas,
a generalization of the
usual QM Pontecorvo formulas.
CP and T violation induced by neutrino oscillations
have also been discussed.

\section*{Acknowledgments}
M.B. and G.V. thank Prof. Y.L.Wu for the kind hospitality at the
``International Conference on Flavor Physics'',
Zhang-Jia-Jie, China, June 2001 and
at the ``International Workshop on Neutrino Physics'',
Beijing, China, June 2001.
This work has been supported by MURST and INFN. Partial support from
 INFM, EPSRC and ESF is also acknowledged.

\end{document}